# RADIATION INDUCED EFFECTS IN A MONOLITHIC ACTIVE PIXEL SENSOR: THE MIMOSA8 CHIP

N. T. Fourches, M. Besancon, R. Chipaux, Y.Li, P.Lutz, F. Orsini

**CEA/DSM/IRFU, CEN Saclay 91191 GIF/YVETTE, FRANCE**

We have studied the effects of ionizing irradiation from a $^{60}$Co source and the effects of neutron irradiation on a Monolithic Active Pixel Sensor Chip (MIMOSA8). A previous report was devoted solely to the neutron-induced effects. We show that extended defects due to the neutron damage changes the distribution of the pixels pedestals. This is mainly due to the increase of the dark generation current due to the presence of deep traps in the depleted zones of the sensors. Alternatively, the exposure to ionizing irradiation increases the pedestals in a more homogeneous way, this coming from the generation of interface states at the Si/SiO$_2$ interface supplemented by the presence of positively charged traps in the oxides. the sensors' leakage current is then increased. We discuss the results in view of increasing the radiation-hardness of the MAPS, bearing in mind that these chips were not designed with any rad-tol layout technique.

## 1- Permanent neutron effects at moderate clocking frequency (1MHz).
### A-Introduction

**a-Hardware and experimental details**

The devices studied here were MIMOSA8 chips developed at CEA/Saclay and IPHC/Strasbourg and manufactured by TSMC through MOSIS. They are based on the pixel architecture, which comprises in-situ amplification and correlated double sampling. Auto-zero discrimination is implemented at the bottom of some columns for the digital outputs that were not used in this study. For the eight analog outputs the useful signal is the difference between a reference signal (calibration) and the readout signal (read) and is computed by software. The DAQ card was developed at IPHC/LEPSI (Strasbourg). The analog signal is measured by an ADC located inside the DAQ board abutted at the output of the measurement board (the so called proximity PCB). Seven chips were irradiated at four different integrated fluxes. Five chips were mounted and measured. The chip N°2 was found defective and not fully evaluated. The last four were extensively characterized. A few un-irradiated reference chips were measured as references. In this work the devices have been measured at a 1 MHz clocking frequency up to $1.13 \times 10^{13}$ neutrons/cm$^2$. The neutron flux has a spectrum which extends up to 20 MeV [1], as indicated in Figure 1.

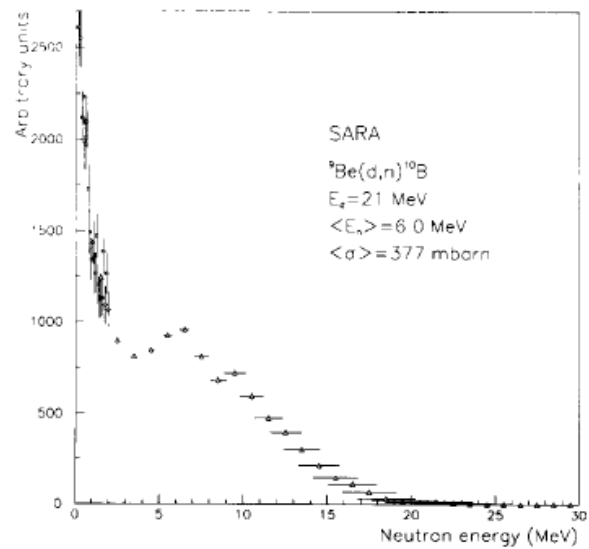

Figure 1: Neutron flux as a function of energy for the source used (J. Collot et al., this figure (Fig 4) is reprinted from J. Collot et al., Nucl. Inst. and Meth. In Physics Research, A 350, 525-529 (1994))

**b-Structure of the pixel diode and hypothetical origin of neutron induced bulk effects**

The pixels designed have diodes of increasing sizes: S2, S3, S4 .S2,S3,S4 stand for S2 subarray, S3 subarray,S4 subarray.

In epitaxial processes, such as the one used for the MIMOSA8 prototype (TSMC 0.25 µm) the degradation due to neutron irradiation is mainly due atomic displacements in the thin epitaxial layer (a few µm). A number of point defects and defect clusters are created; this reduces the lifetime of the minority carriers (in these case electrons) and induces deep traps that affect the transport properties in the silicon.

There are clearly two kinds of effects:

- A continuous effect is the increase of the generation-recombination current in the depleted region of the n+p junction of the sensing element. The sensing node integrates this current and consequently acts as a probe. This current is directly related to the electron lifetime and is normally enhanced when the neutron exposure is increased.

- Carrier capture affects the free drift length of the electrons and their diffusion length. In the case of the MAPS (Monolithic Active Pixel Sensors) electron transport also occurs by diffusion as the applied electric field in the sensor is extremely low. If $\sigma_n$ is the capture cross section for electrons, the free drift length is deduced from: $N_t$ v $\sigma_n$ where $N_t$ is the deep trap concentration. For one type of trap only, the free drift length is then: $L=1/(N_t\sigma_n)$. L is reduced when the trap concentration increases, along with the neutron integrated flux. A similar conclusion can be drawn with the diffusion length which is given by: $L = (D_n\tau_n)^{0.5}$. The lifetime is reduced when the trap concentration increases and this induces a reduction of the diffusion length. The lifetime is related to the trap concentration (for identical traps) by: $\tau_n \sim (vN_t\sigma_n)^{-1}$.

**B-Measurements results and qualitative interpretation**
**a- Effects on the pedestals:**

For the MIMOSA8 chip the analog outputs are processed in order to subtract the voltage levels present before and after reset. This enables the extraction of the useful signal and a reduction of the pedestals. The residual pedestals (offset) are related to the characteristics of the sensing elements and the material. Figure 2 (a) (b) shows the distribution of pedestals on four columns of the array for a sample irradiated at the maximum integrated flux and at the minimum integrated flux. Figure 3 and Figure 4 show the variation of the pedestals with integrated flux for the central columns and the lateral columns.

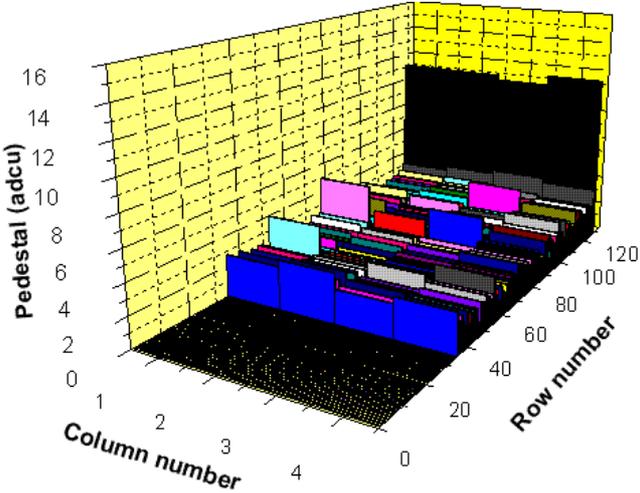

Figure 2 (a): distribution of pedestals on a subarray of four columns (central) for a chip irradiated at $1.44\ 10^{11}$ neutrons/cm$^2$.

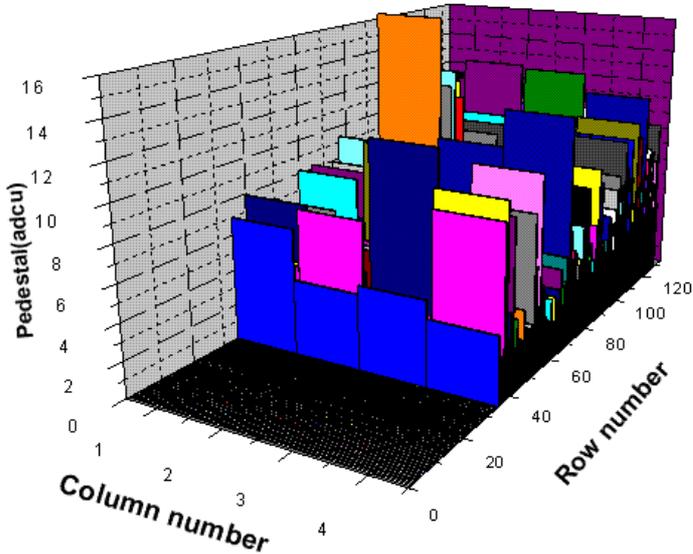

Figure 2 (b): distribution of pedestals on a subarray of four columns (central) for a chip irradiated at $1.13\ 10^{13}$ neutrons/cm$^2$.

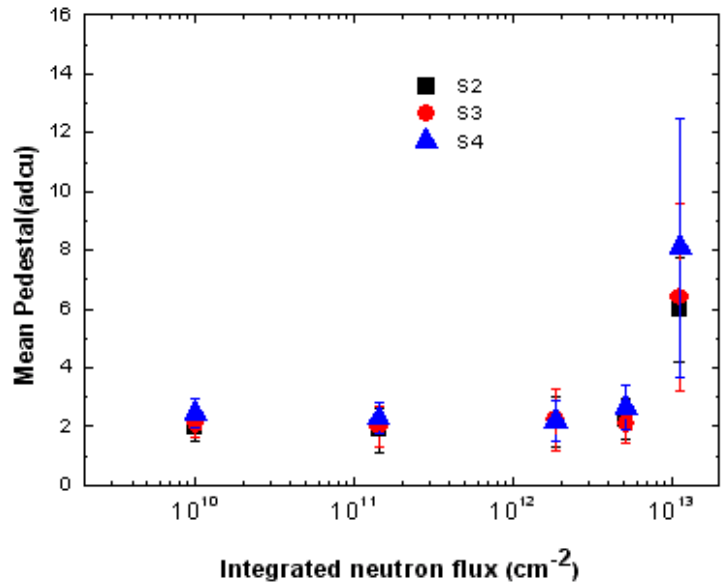

Figure 3: (mean) pedestals versus neutron-integrated flux for the central columns. All the four columns were taken; the first point (on the left side of the plot) corresponds to a non-irradiated sample. The error bars correspond to the sigma on the pedestals.

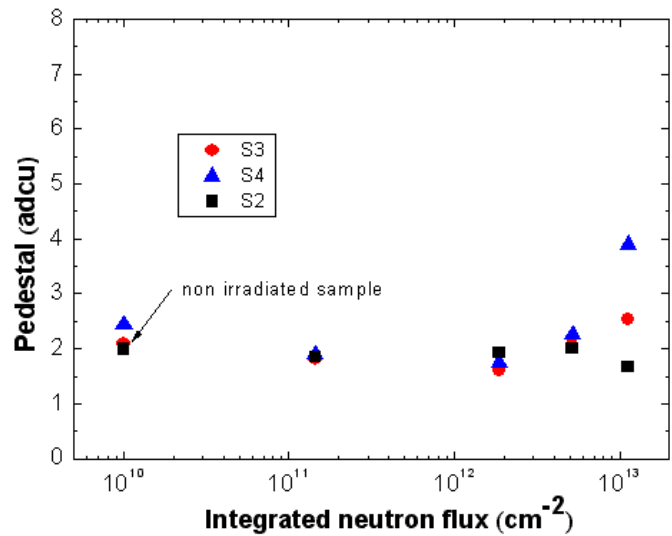

Figure 4: pedestals versus neutron integrated flux for the lateral columns. All the four columns were taken; the first point (left) corresponds to a non-irradiated sample. No errors bars are displayed here.

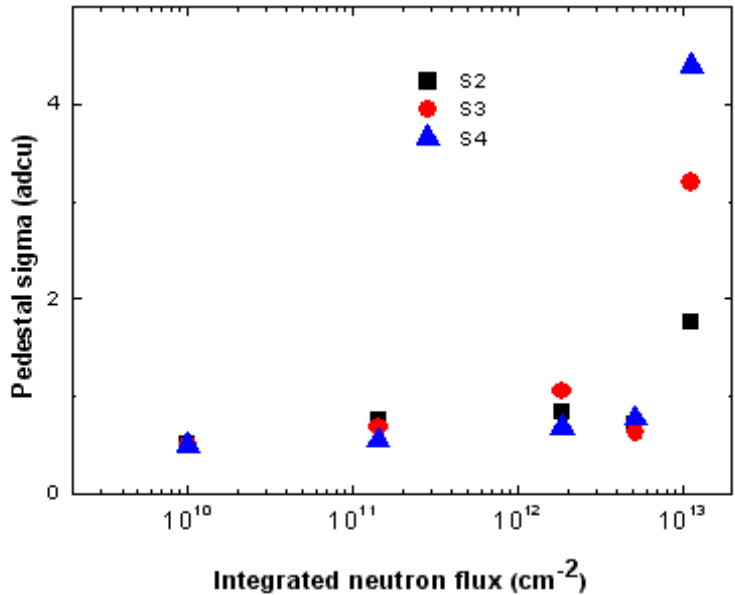

Figure 5: pedestal sigma (fixed pattern noise) versus integrated neutron flux for the 3 sub-arrays studied. All the four central columns were taken in the computation.

    Two conclusions can be drawn. First, the pedestals mean value increase with total integrated neutron flux. Second, the dispersion between pedestals increases with the total integrated flux. This behaviour does not depend on the column as it can be seen on Figure 2. The two plots seem to show that for the Fixed Pattern Noise (Pedestal Sigma) varies more significantly than the pedestals. The pedestals alone vary by less than a factor of 3.
    The dispersion between neighbouring pixels (as it can be observed in Figure 2) may find its origin in a non-homogeneous distribution of neutron-induced defects in the sensitive silicon epitaxial layer. ).The dispersion of the pedestals (the fixed pattern noise) increases with the integrated neutron flux (Figure 5). It seems clear clear that a sharp increase in the FPN occurs for the highest integrated fluxes. This might be related to the distribution of neutron-induced clusters, as it will be discussed next.

**b-Temporal noise:**

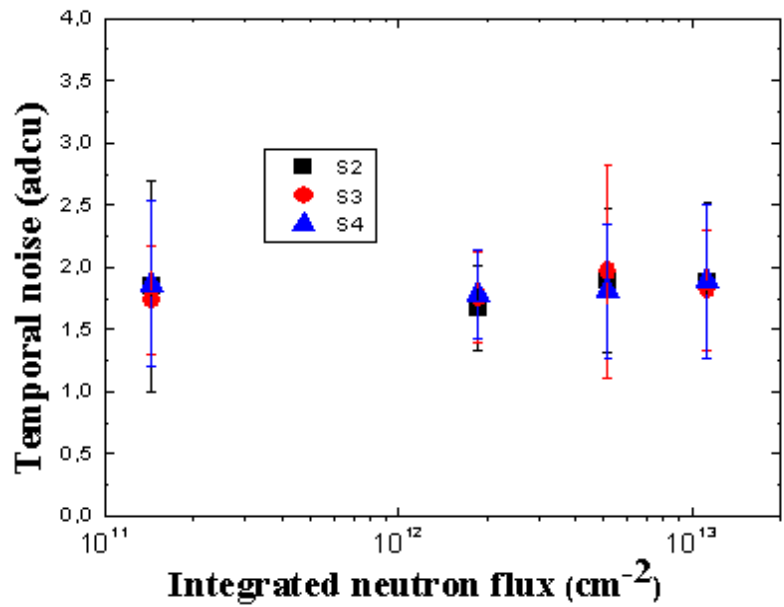

Figure 6: Temporal noise versus integrated neutron flux in adc units for the three sub arrays.

The temporal noise of the three sub-arrays is weakly dependent on the total integrated flux. There is no perceptible temporal noise increase.

Flicker noise due the increased concentration of radiation-induced traps should be a major source of bulk detector noise. But the noise from the readout electronic should be dominant before irradiation. At this stage of the study it is difficult to assess the respective contributions of these noise sources (Figure 6). A simple analysis of the data shows that most of the noise should be due to the readout electronic

**c- Charge collection efficiency:**

The charge collection efficiency (CCE) is here defined as the ratio of the position (Most Probable Value of the signal amplitude) of the peak corresponding to 9 pixels clusters to the calibration peak (total collection of the charge). The clusters are here calculated in a 4 columns to 32 rows sub-array. The charge collection efficiency is directly related to the electron transport properties in the silicon.

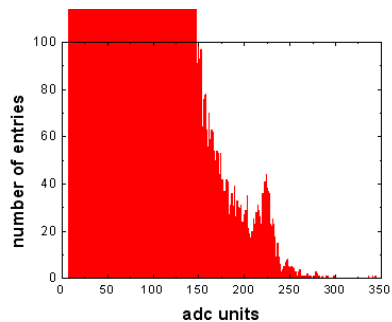

Figure 7(a): calibration (total absorption) peak for the lowest integrated flux (1 adcu=0.5 mV),S2 sub-array.

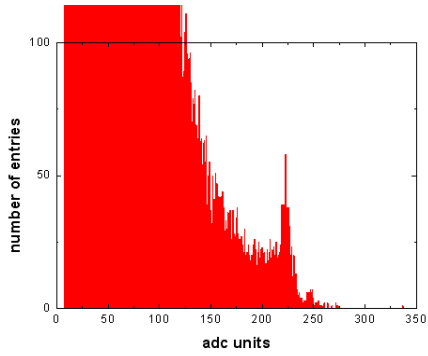

Figure 7(b): calibration (total absorption) peak for the highest integrated flux (1 adcu=0.5 mV), S2 sub-array.

Experimentally the calibration peak shape does not change greatly with integrated flux (Figure 7(a) (b)). The cluster peak shifts towards lower values (Figure 8 for the highest irradiation).

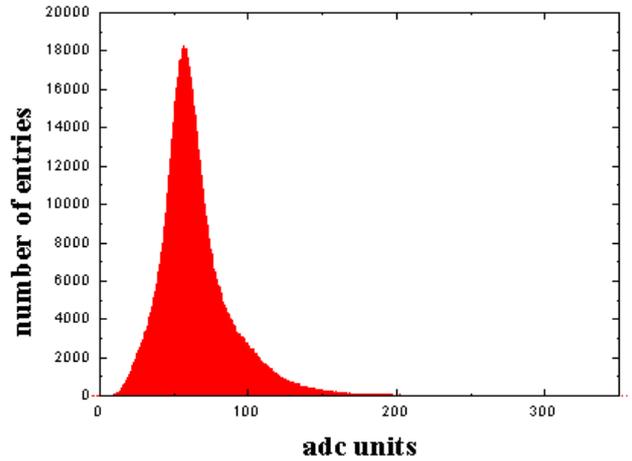

Figure 8: cluster peak (9 pixels) for the highest integrated flux (1 adcu=0.5 mV), S2 sub-array.

The charge collection efficiency plotted versus integrated neutron flux is given in Figure 9.

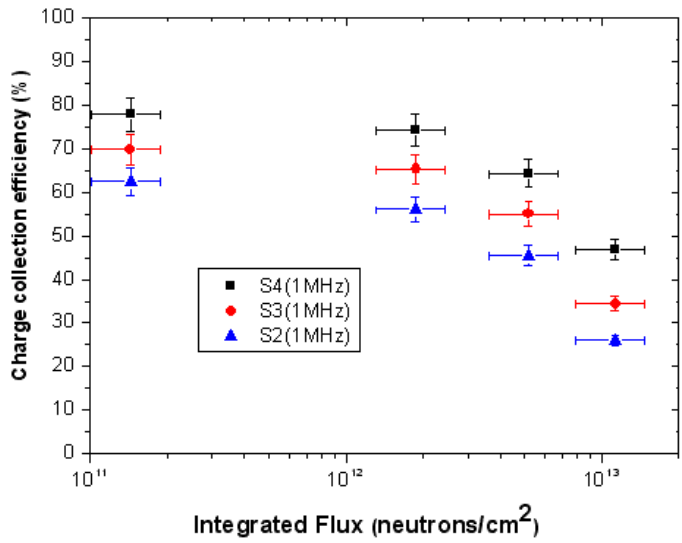

Figure 9 (a): charge collection efficiency versus integrated neutron flux for the three sub-arrays studied (log scales).

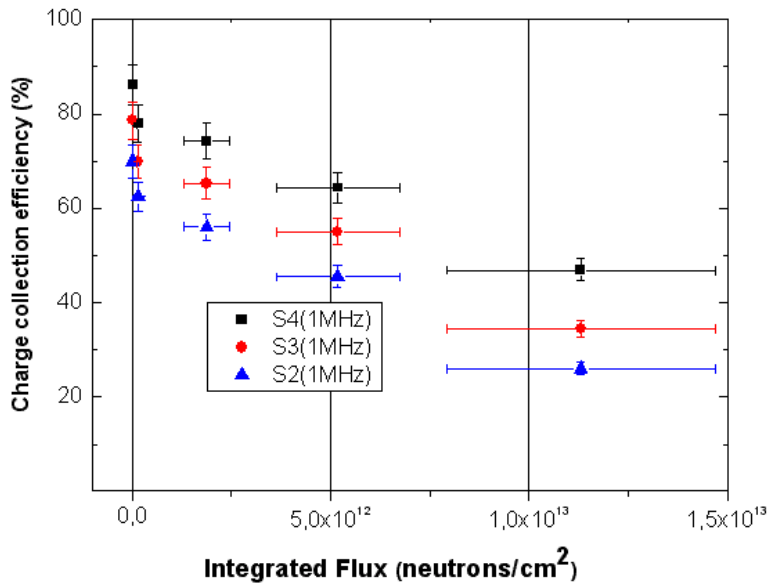

Figure 9 (b): charge collection efficiency versus integrated neutron flux for the three sub-arrays studied (linear scales). The point at the origin is that of an un-irradiated chip.

The absence of the left side shouldering peak in Figure 8 could find its origin in the low intensity of partial epi-layer events which cannot reach the S/N threshold due to charge lost in radiation induced traps, supplemented by the shift of the main peak towards low adc values.

It is clear that the CCE decreases markedly for integrated fluxes above $10^{12}$ neutrons/cm$^2$. In this case as the pixel clusters have an area nine times the area of one single pixel, the sensitivity to neutron irradiation is increased. Therefore the Charge Collection Efficiency is more affected by irradiation than the pedestals.

Deep traps reduce the average electron-migration length and lower the average signal amplitude through various mechanisms.

In addition, Figure 10 shows the area of the calibration peak versus the integrated neutron flux for the S2 sub-array. A trend appears which could suggest that the number of calibration events slightly decreases when the integrated flux increases. This may be understandable: the calibration peak is made up by rare events (hits) that result in an almost total collection of the generated carriers (electrons). As the concentration of radiation induced defects increases the number of carriers collected with no capture decreases significantly. However this has proved to be difficult to check thoroughly.

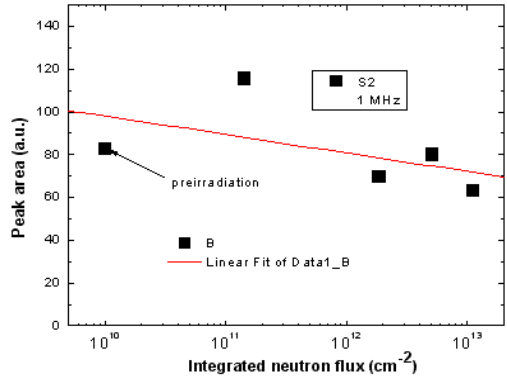

.
(a) area of the pedestal component of the calibration peak

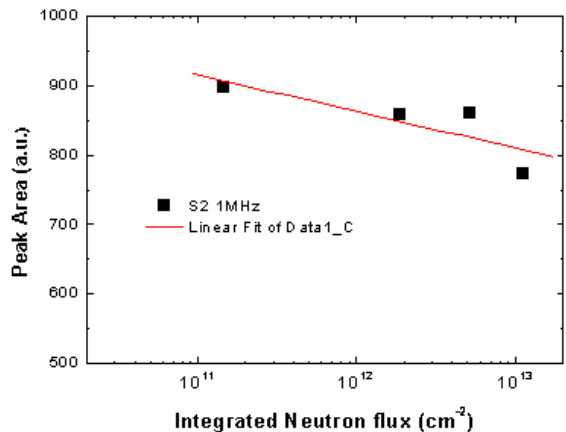

(b) area of the gaussian component of the calibration peak minus pedestal

Figure 10 (a) (b): area of the calibration peak versus integrated flux (S2 sub-array) with a linear fit. The peak area is deduced from a fit based on a Gaussian distribution supplemented with a residual pedestal.

## 2- Experimental evidence for single pixel effects

Neutron irradiation should affect single pixels but not uniformly change the characteristics of the pixel array. This is verified for the pedestals that are increased in absolute value by neutron irradiation and ionizing irradiation. However the results are very much different. Figure 11 show an image of the pedestals both for a fresh and a neutron irradiated chip.

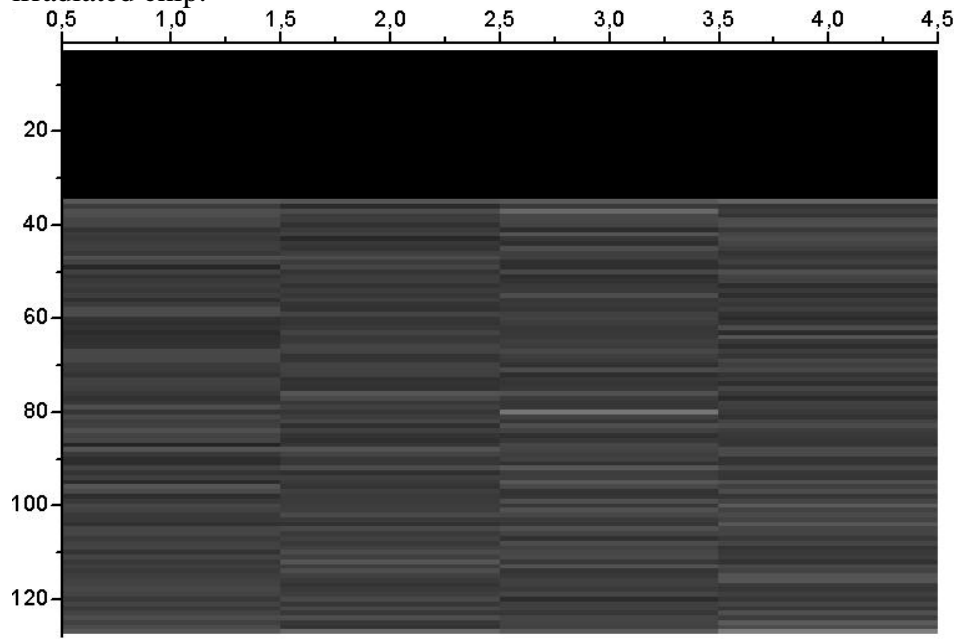

Figure 11 (a): image of the pedestals of an low fluence irradiated chip(0:dark to 8:bright adcu)

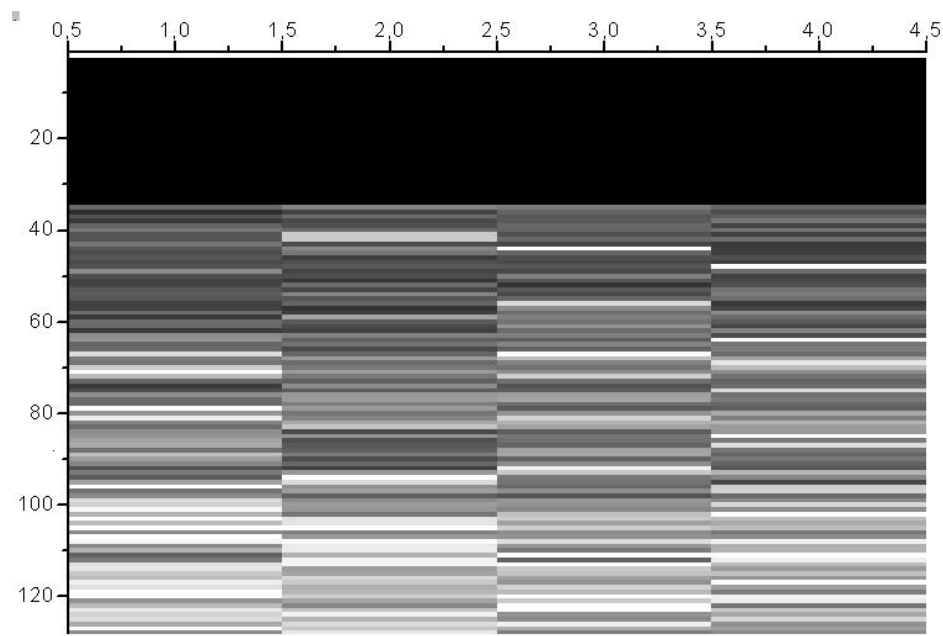

Figure 11 (b): image of the pedestals of an high fluence irradiated chip (0:dark to 8:bright adcu)

It is clear that white (high pedestal) pixels are often isolated and that the effect is not uniform as stated earlier. There seems to exist an moderate frequency spatial noise in neutron irradiated chips that does not exist in undamaged chips. The effects of ionizing irradiation are closer to uniformity as recalled in Figure 12. This due to the nature of neutron-induced defects that create a localized damaged area below the pixel surface that does not extend more than a few thenth of µm in linear range thus lower than the size (pitch) of a pixel.

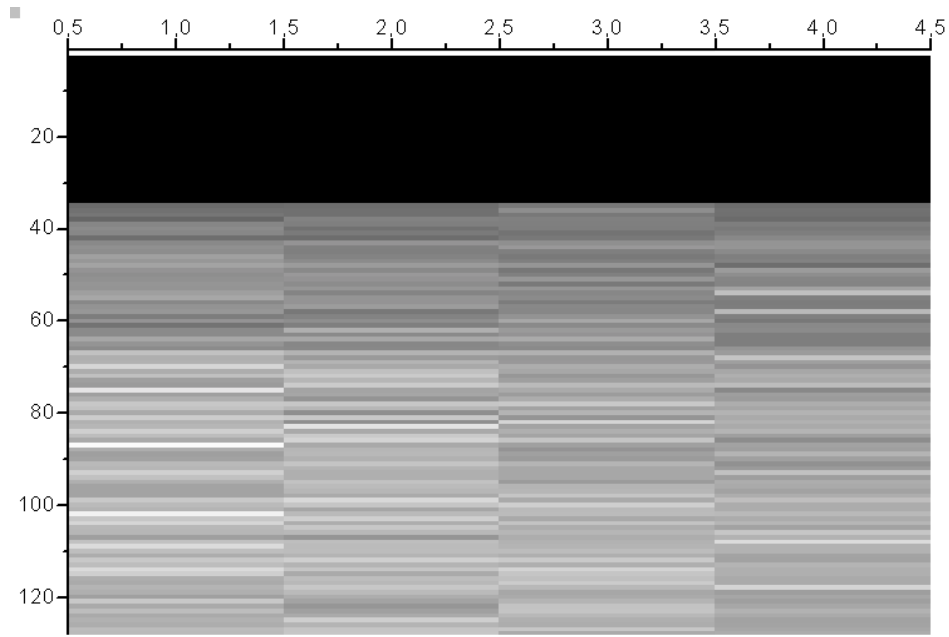

Figure 12: same image on a 137 krads irradiated chip (not the same tone scale 0: dark , 100 bright, adcu).

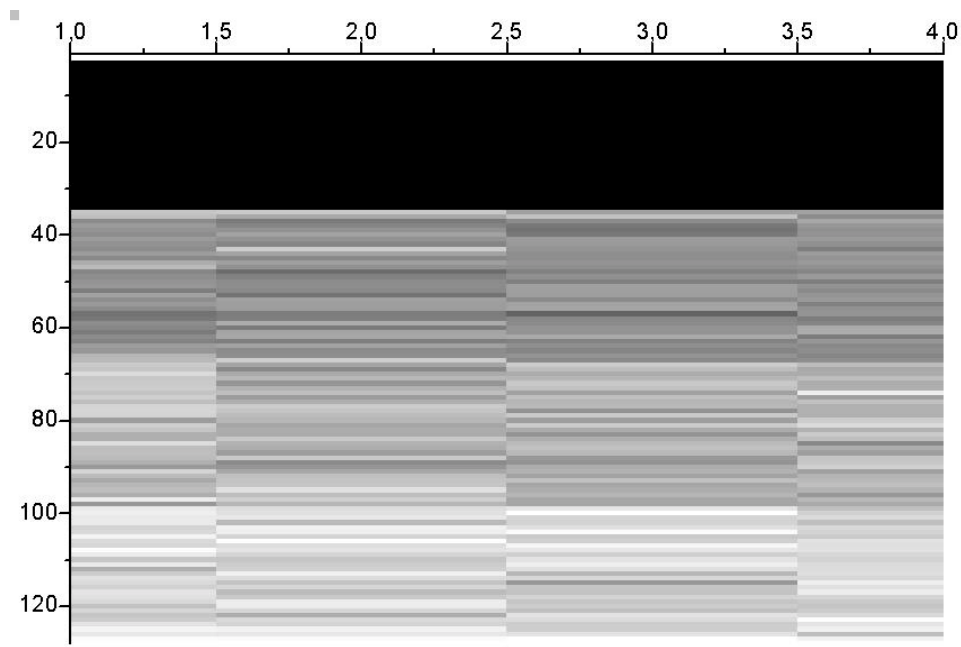

Figure 13: same image on a 34 krads irradiated chip (scale 0: dark, 8 : bright, adcu).

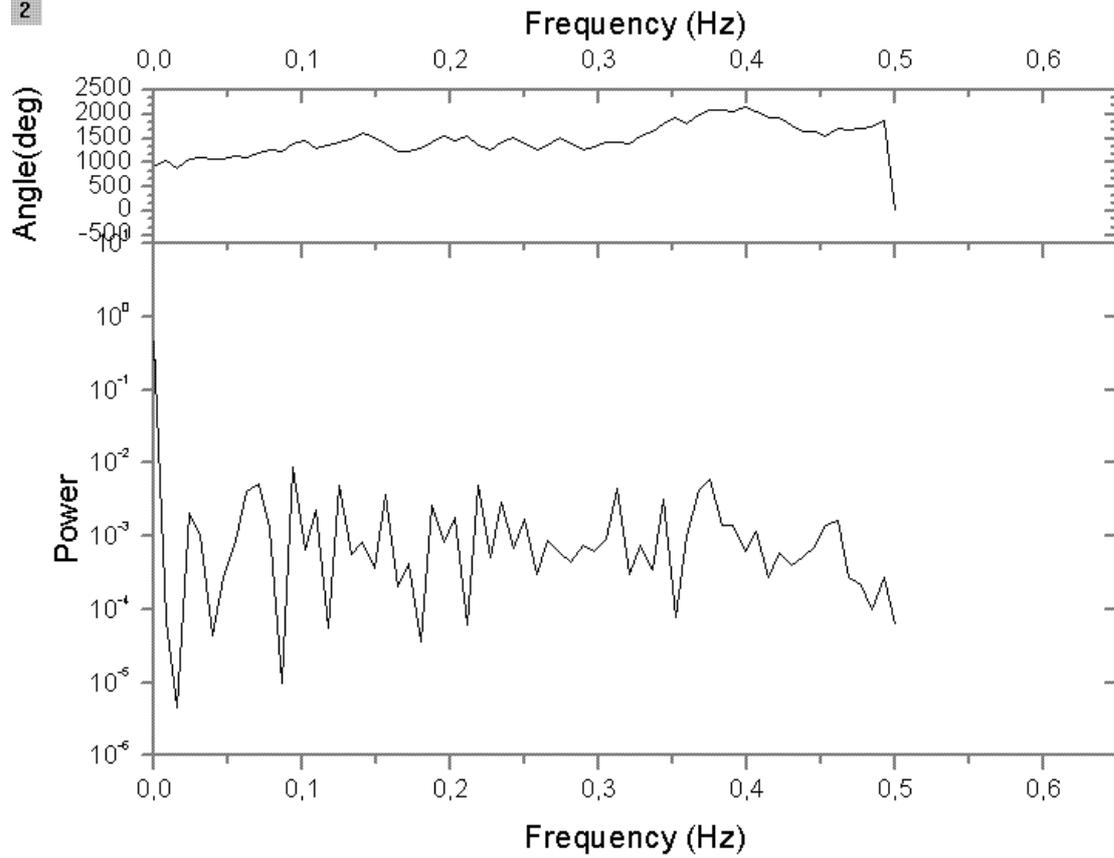

Figure 14 (a) : one dimentional FFT on one high fluence neutron-irradiated chip, the frequency is spatial

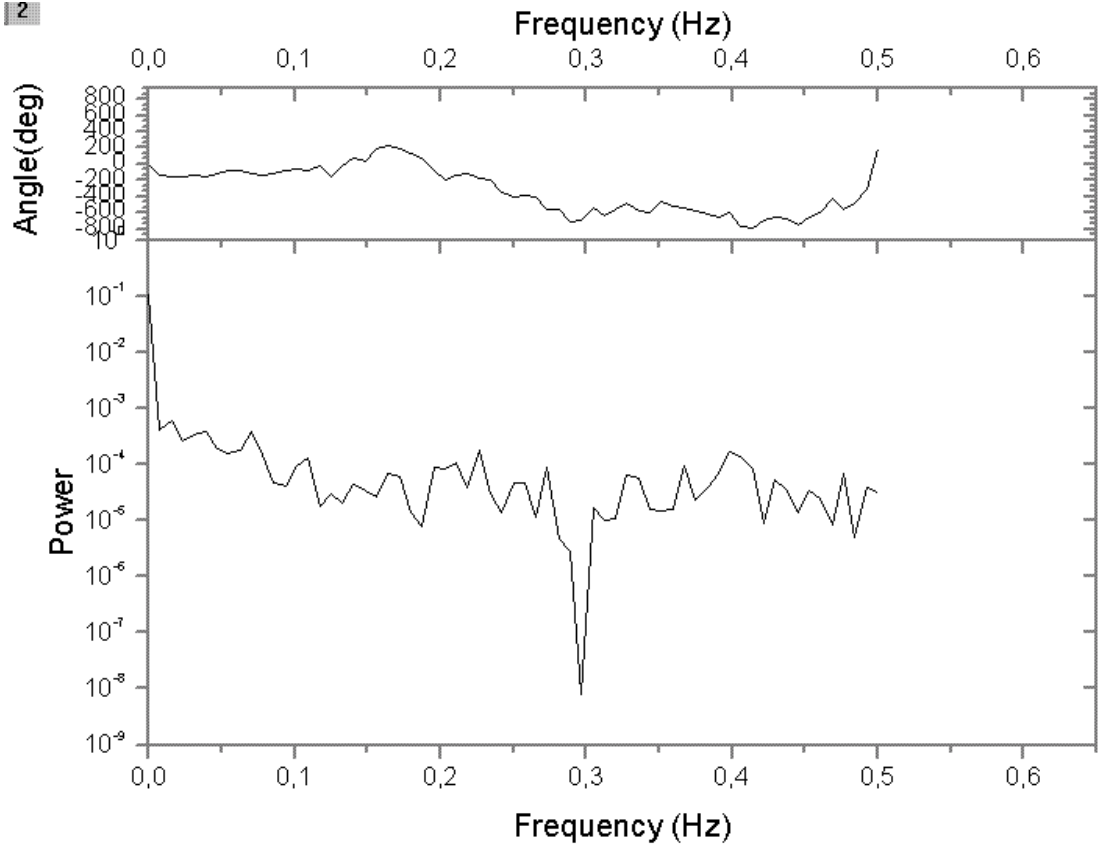

Figure 14 (b) : one dimentional FFT on one low fluence neutron irradiated chip, the frequency is spatial

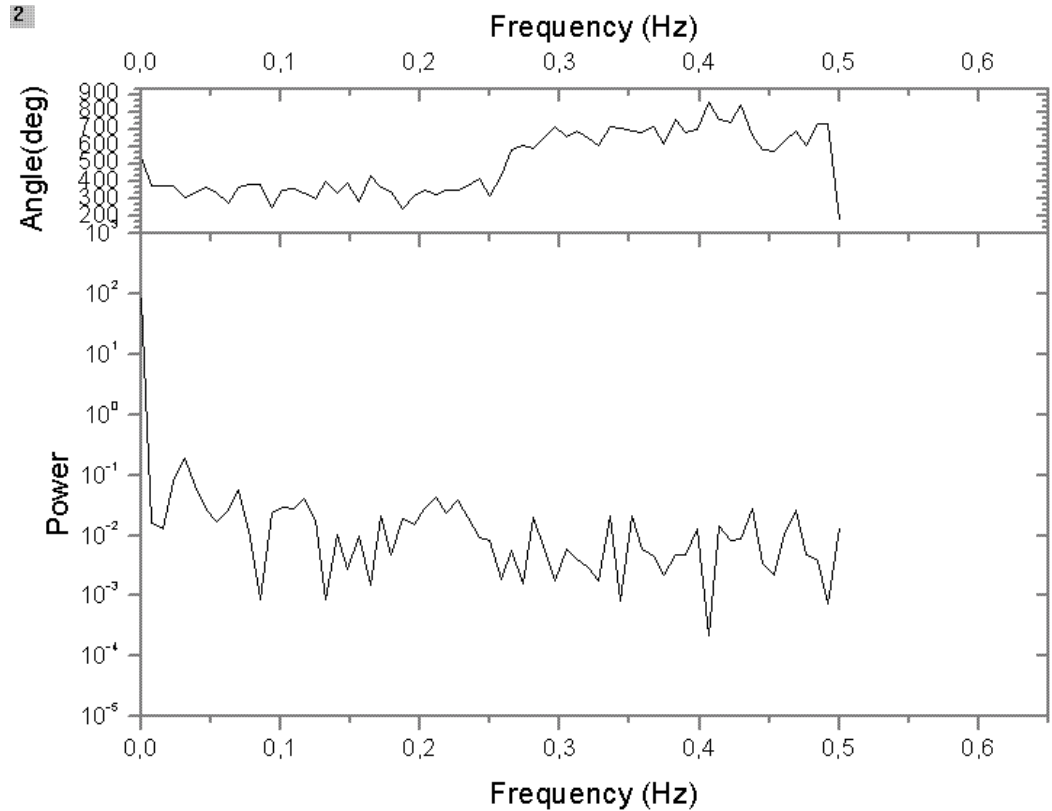

Figure 14 (c): one dimentional FFT on one photon irradiated (137 krads) chip, the frequency is spatial.

FFT can be useful to quantify this effect. Figure 14 show the one dimentional FFT on unirradiated, neutron irradiated and photon (ionizing) irradiated chips. There is always the presence of a similar low frequency component. These plots show that the shape of the spectral density is not significantly altered by ionizing irradiation. For the high fluence neutron irradiated chip the power spectral density is flat and remains at a high level, the low frequency component vanishes. These results complement the previous analysis made for single pixel effects on the pedestals.

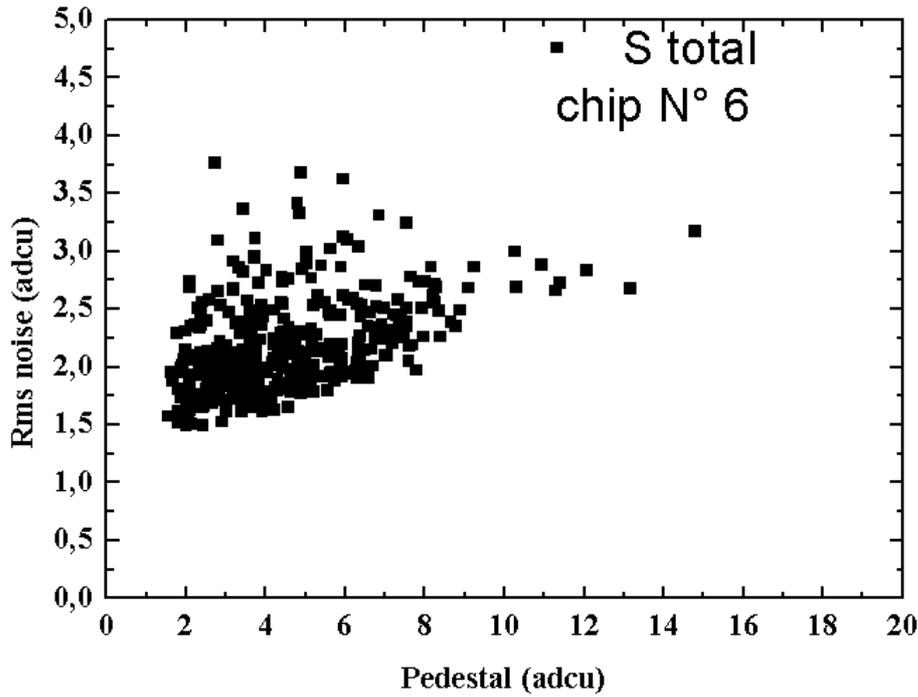

Figure 15: plot of the temporal noise value versus pedestals value in adc units on an high fluence ($10^{13}$ cm$^{-2}$) irradiated chip.

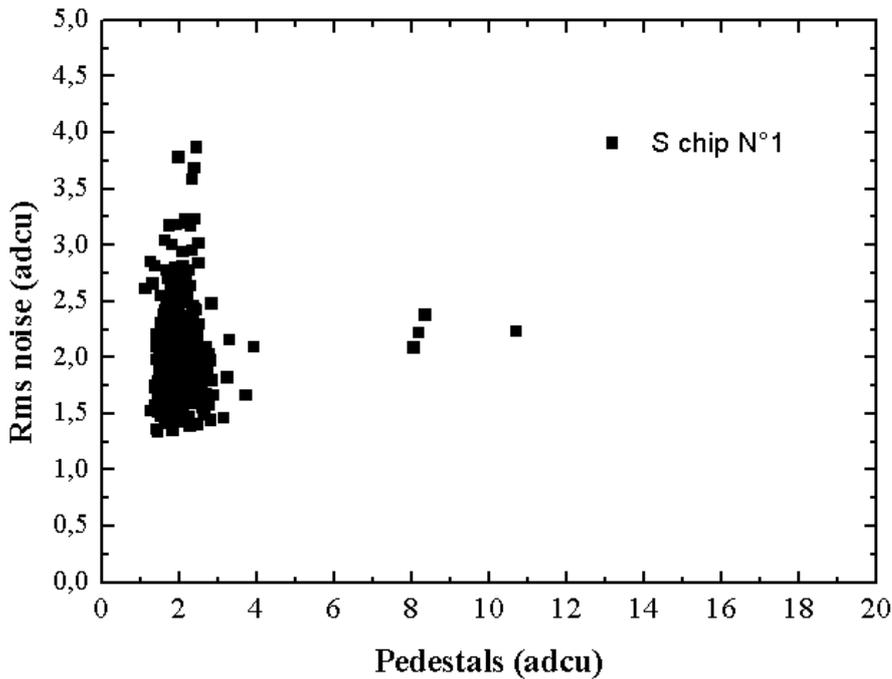

Figure 16: plot of the pedestals versus temporal noise in adc units on an low fluence (~ $10^{11}$ cm$^{-2}$) irradiated chip

Figure 15 shows that the pixels with high pedestals have high temporal noise on all the sub-arrays. Pixels that exhibit high pedestal increase have high rms noise increase. It is clear from Figure 16 that the scatter in temporal noise is important but the width of the pedestal distribution is then reduced.

# 3-Permanent Ionizing Radiation Induced Effects at 1 MHz Clocking Frequency

## A- Effects of clocking frequency

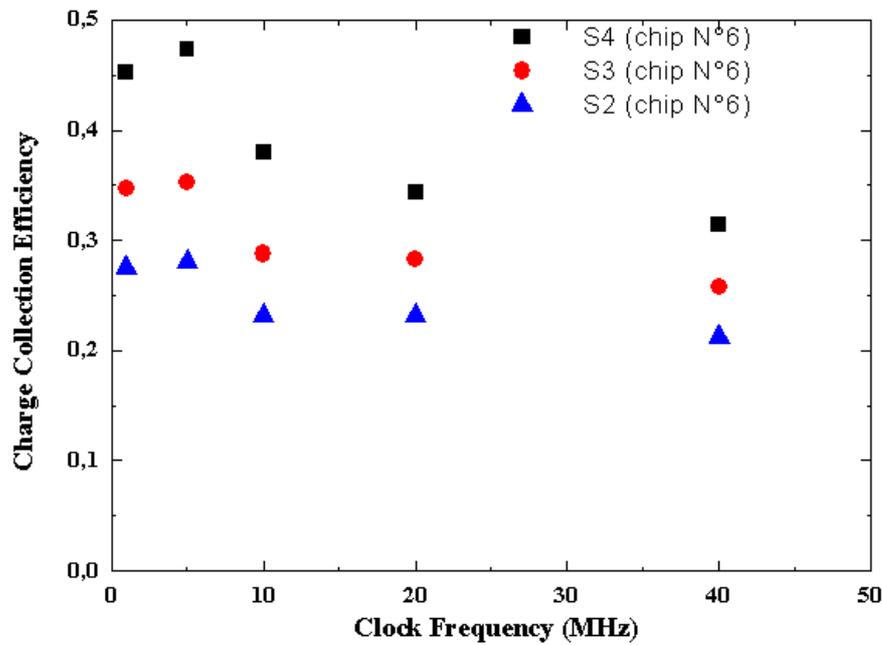

Figure 14: charge collection efficiency versus clock frequency for the most irradiated chip.

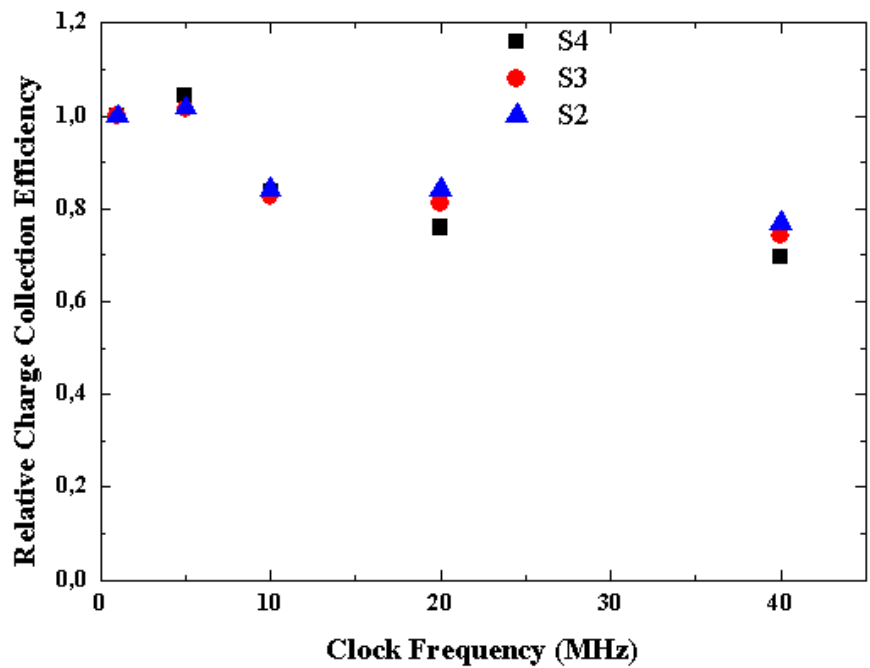

Figure 15: normalized (to the 1 MHz value) CCE versus clocking frequency for the most irradiated chip.

The charge collection efficiency lowers as the clocking frequency increases but remains at reasonable values (Figure 14). A drop of around 20 % is observed at 40 MHz (Figure 15). The behaviour of the CCE is similar for all the sub-arrays studied. Measurements were made up to 40 MHz but the functionality of un-irradiated chips is kept up to more than 100 MHz as shown in earlier studies.

## B- Results for ionizing doses up to 280 krads and qualitative interpretation

Irradiations were performed at room temperature on a single chip measured before each irradiation stage. Doses were progressively increased and the cumulated doses reached approximately 280 krads. The conversion factors, the pedestals, the charge collection efficiency are plotted versus ionizing dose. Except for the pedestals the characteristics were not easily evaluated above 137 krads dur to the degradation of the chip.

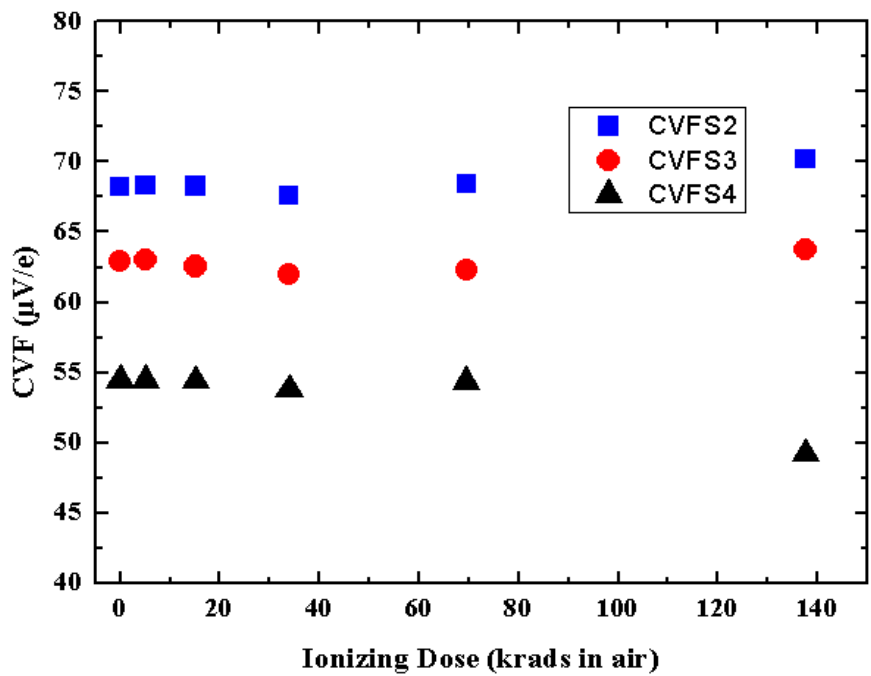

Figure 16: Conversion factors for the three sub-arrays versus integrated ionizing dose.

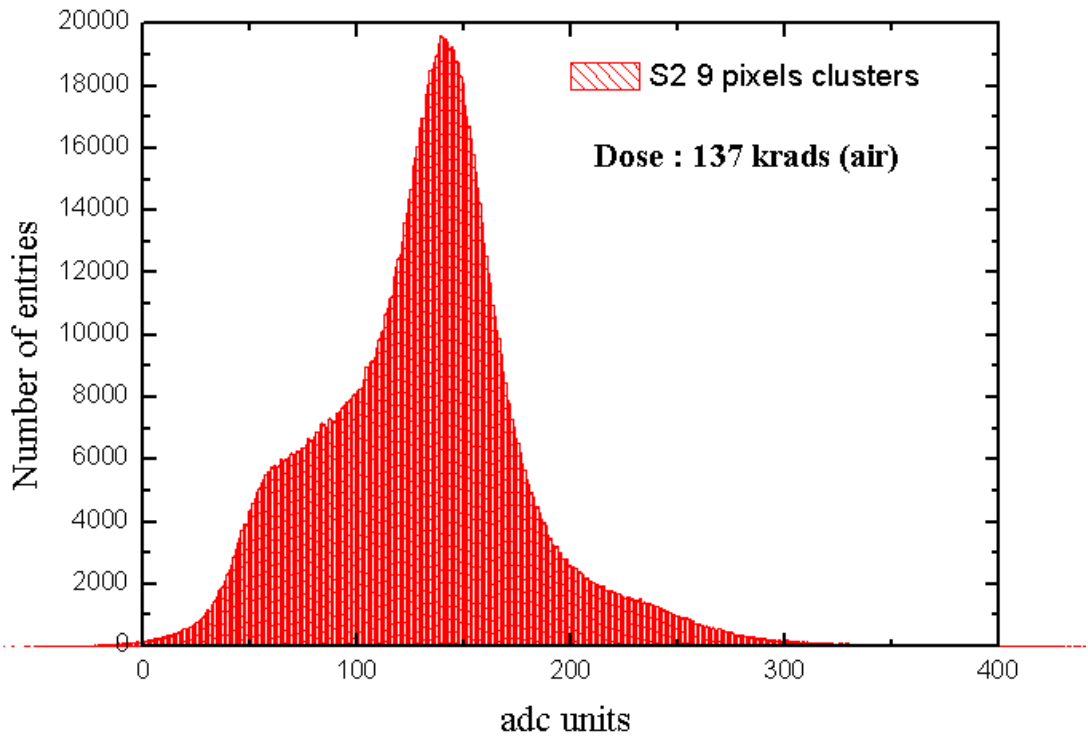

Figure 17: 9 pixels cluster spectra for the $^{55}$Fe photon source.

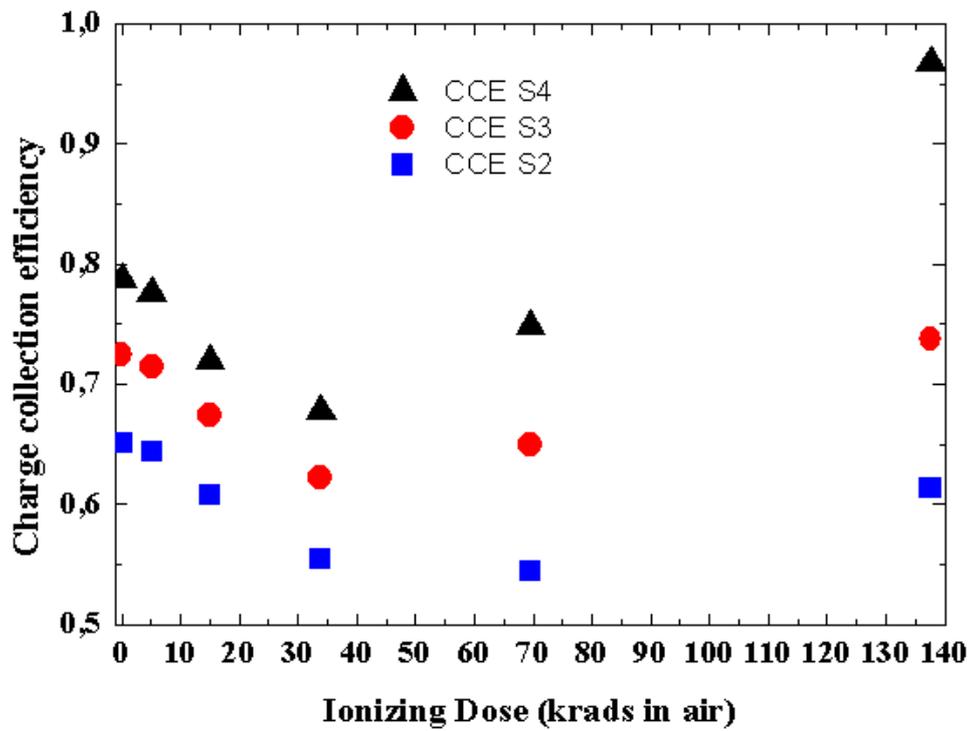

Figure 18 : charge collection efficiency of the three sub-arrays plotted versus ionizing dose at 1 MHz clocking frequency.

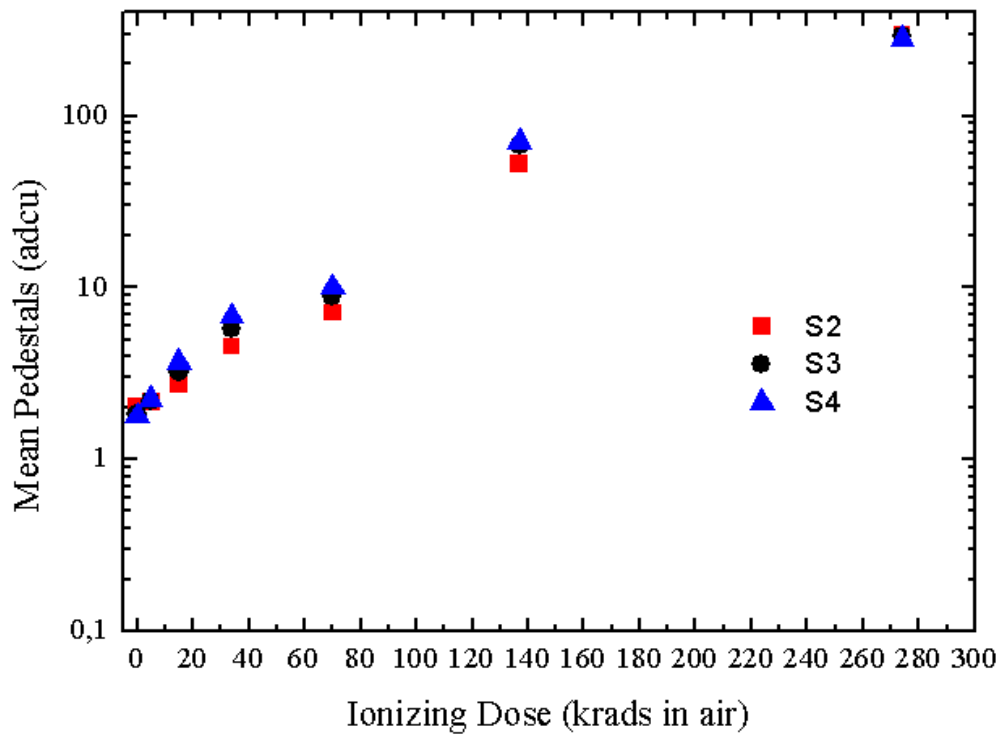

Figure 19 : Mean pedestals versus integrated ionizing dose for the three arrays studied.

The CVFs' (Figure 16) are remarkedly stable with integrated dose and remain similar to the one measured on unirradiated chips. This proves that for all the sub-arrays the gain of the electronic stages are constant and that the MOS amplifying and processing circuits are tolerant to radiation up to 140 krads. The drop in charge collection efficiency up to medium doses (Figure 17-18) can be interpreted by the effect of negatively charged interface states which modify the electron collection effectiveness of the upper electrodes. In addition interface recombination may occur. The CCE increase at higher doses should be due to the radiation generated positive charge in the thick oxide that enhances the drift of event induced electrons towards the upper electrode. This effect balances the action of interface negative charges.

The pedestals are increased in averaged with irradiation dose (Figure 19).The growth is slightly sublinear and very close for each sub-array. One major cause can be pointed out. As well as in neutron irradiation experiments the rise of the leakage current of the sensing element is the main reason for the pedestal increase. The leakage current should be here dominated by a near surface current due to the thick oxide interface states of a volume leakage current in the thin oxide of the MOS readout or reset transistors. Both depend on the concentration of radiation induced defects in the $SiO_2$ volume or at the $SiO_2/Si$ interface. They are physically similar to recombination/generation currents.

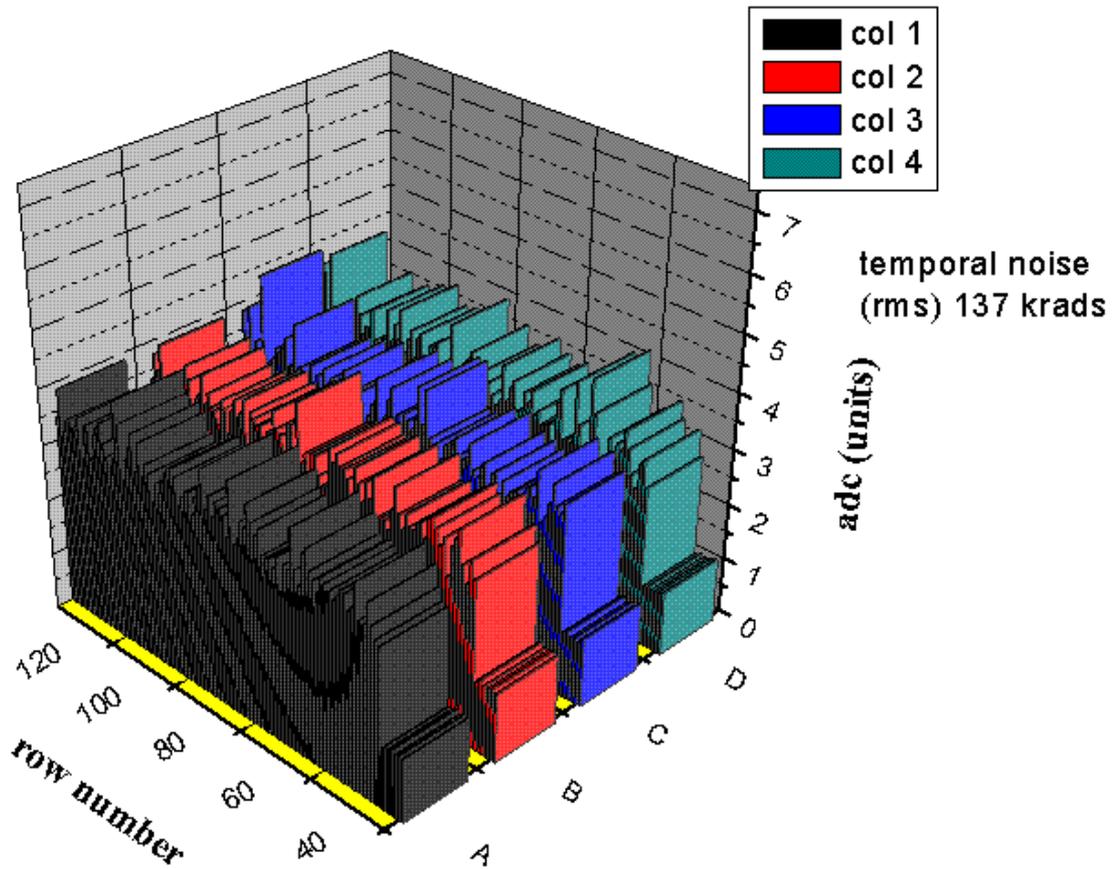

Figure 20 : temporal noise of the three sub-arrays for a 137 krads integrated dose.

The temporal noise is slightly increased after irradiation. This could be the effect of the MOS transistors degradation, probably flicker noise increase due to oxide trap generation by irradiation. However sensor diode leakage current increase could be one contributor to this rise for large enough doses.

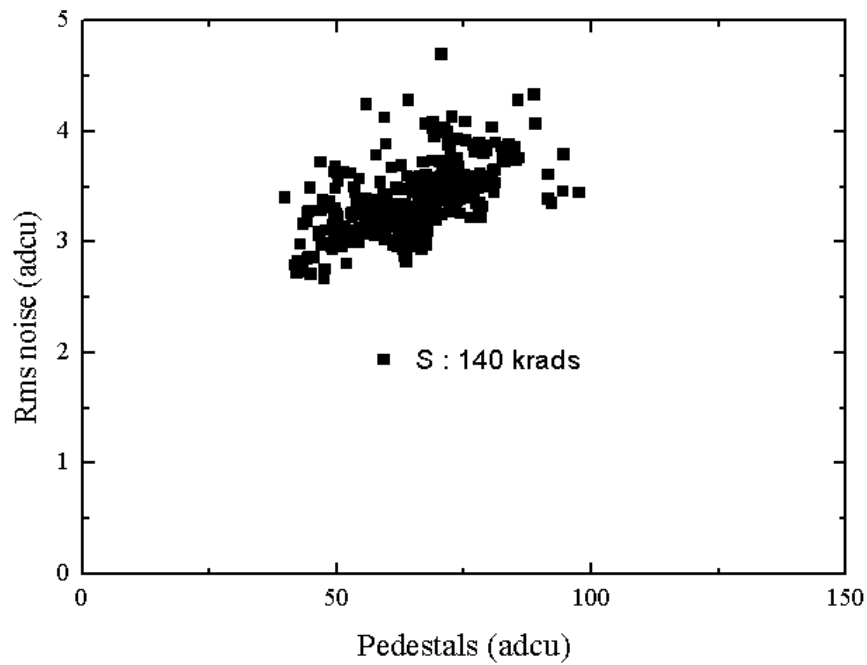

Figure 21 : distribution of temporal noise values versus pedestal values for a 137 (140 targeted) krads irradiated array

The temporal noise remains more stable than the pedestals after irradiation but pixels with large pedestals are noisier than the others (Figure 22). The characteristics of a fitting straight line differ to those obtained on neutron-irradiated samples, suggesting different degradation mechanisms (Figure 22).

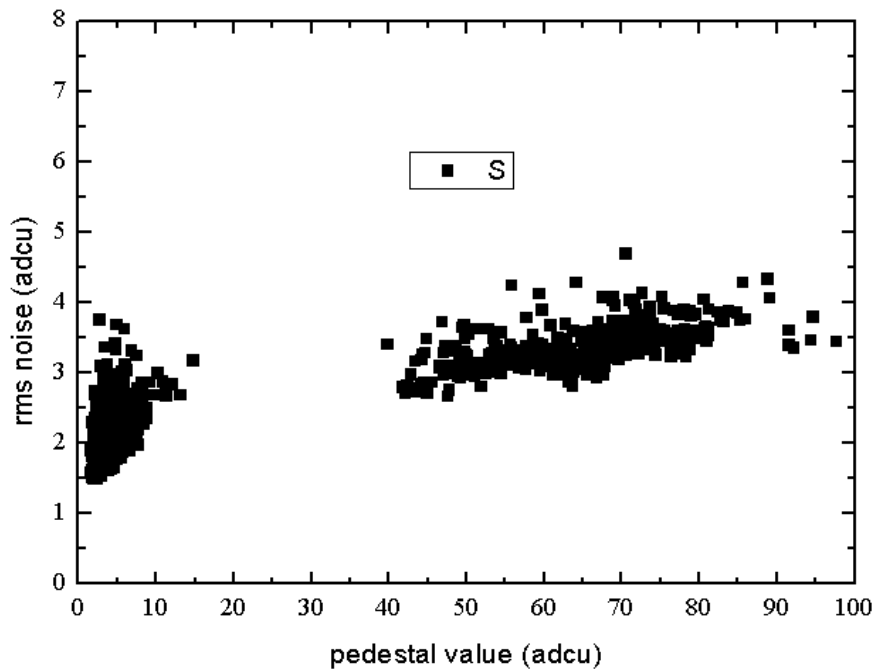

Figure 22: comparison between the neutron and ionizing irradiated chips , noise and pedestals

# 4-Discussion

## A-Overall description:

Previous work shows that irradiation by massive particles in semiconductors produce defects that trend to form clusters of significant spatial extension up to 1 µm [2][3][4]. The collision cross section for 1 MeV neutrons on silicon is approximately $3 \cdot 10^{-24}$ cm$^2$ [5]. Simple calculations show that for an integrated flux of $10^{13}$ neutrons/cm$^2$ the total number of atomic collisions per squared centimetre is of the order of $10^8$ collisions /cm$^2$, assuming a 10 µm epitaxial layer thickness. For a 25 µm large pixel the number of primary collisions in the epi-layer is roughly 5. For MeV energetic neutrons each primary collision induces a cascade of approximately 500 secondary displacements. The threshold energy for atomic displacements in silicon is 20 eV and the energy of the Primary Knock on Atom (PKA) is roughly 36 keV on average, leading to around 500 atomic displacements (for the principle of the calculation see [6] and [7]). As a result, the number of primary point defects in the active region of the pixel lies near 2500 at $10^{13}$ neutrons/cm$^2$. This relatively low value may explain the important dispersion of the pedestal values after irradiation. For integrated fluxes 100 times lower ($10^{11}$ neutrons/cm$^2$), the number of defects below each pixel has an average lower than 5. This indicates that the effect of neutron irradiation at this integrated flux is very low as it can be seen in Figure 2(a). Moreover as the number of primary collisions per pixel is close to unity even at high integrated fluxes the spatial extension of the defects (1 µm in average) results in defect clusters non-homogeneously distributed in the pixel inducing a possible dispersion of leakage currents in the sensor. This is the basic explanation of the dispersion of the pedestals through the array. Images of the pixels' pedestals in the arrays tend to support this idea. In addition the number of primary collisions in a cluster of 9 pixels is 45 for $10^{13}$ neutrons/cm$^2$. This may indicate that for integrated fluxes lower than $2 \cdot 10^{11}$ neutrons/cm$^2$ the effects of neutron irradiation should be reduced to weak single pixel effects. This could to be the case Moreover the dispersion of these pedestals should saturate if the total irradiation defect concentration is increased above a threshold. This should however be confirmed experimentally.

## B- Leakage current: interpretation

For a 1 MHz clocking frequency the time interval between two readouts is 128 x 1µs x 16 = 2.048 ms (rigorously speaking 2.048 ms – 16 µs). When no events occur this is the time interval between which the leakage current is effective in changing the pedestals. For the highest integrated neutron flux and the largest diode (S4) the pedestal value is around 8 adcu, corresponding to 4 mV. This means that the slew rate is given by: dV/dt = 4 mV/2.032ms = 1.97 V/s. For the S4 sub-array the conversion factor is 55 µV/e-, equivalent to $3.42 \cdot 10^{14}$ Volt/Coulomb. The equivalent capacitance at the sensing node is then $2.92 \cdot 10^{-15}$ F or 2.92 fF. The leakage current is then given by: i=C(dV/dt) = $1.97 \times 2.92 \cdot 10^{-15}$= $5.75 \cdot 10^{-15}$ A=5.75 fA. This value is relatively weak and proves that the recombination-generation current remains low.

$J= qW <N_t> \sigma_p \sigma_n v_{th} n_i / (\sigma_n + \sigma_p)$ is the current density in the diode due to recombination processes.
With W~1µm, order of magnitude of the depletion region below the n+ région of the sensing

diode:

$\sigma_n = \sigma_p = 10^{-15}$ cm$^2$. $v_{th} \sim 10^7$ cm.s$^{-1}$, $n_i \sim 1.45 \times 10^{10}$ cm$^{-3}$.

$J \sim <N_t> 1.17 \times 10^{-21}$ A/cm$^2$. The concentration of traps can be computed with a current density of: $J \sim 0.0998 \times 10^{-6}$ A/cm$^2$ assuming squared diode (2.4 µm x 2.4 µm). The trap concentration is then: $<N_t> \sim 8.53 \times 10^{13}$ cm$^{-3}$. This gives an electrically active defect introduction rate of approximately: $K = <N_t>/\Phi = 7.5$ cm$^{-1}$. This value may be compared to the value of 5 cm$^{-1}$ that can be deduced from the numerical estimations introduced in paragraph 5 or to elder experimental values [8].

## C-Comments

At present the results show that for integrated fluxes lower than $10^{12}$ neutrons/cm$^2$ the effects on the pedestals, the CCE and the temporal noise are insignificant. Single pixel effects have been observed and should deserve studies that are more thorough. Quantitative pedestal studies have shown that the epitaxial-silicon material should be improved with regard to its radiation tolerance. However these measurements readily prove that the present characteristics of the MIMOSA8 chips are sufficient to satisfy the needs of the future linear colliders in terms of atomic displacement effects due to neutrons.

# 5- Conclusions

Ionizing irradiation effects have been studied up to 280 krads. As the chip was not designed with radiation tolerance layout rules, functionality was lost between 137 and 280 krads. Results are somehow encouraging with respect to total dose tolerance. Present level neutron induced effects maintain MAPS characteristics at acceptable levels for ILC applications. However, thorough studies are needed to investigate the effects of doping levels in the epi-layer, epi-layer thickness and the nature of the impurities on neutron irradiation tolerance, in the view of increasing the range of applications. A significant ionizing radiation tolerance can be obtained by design efforts (thin oxides substitution, closed shape transistors,guard rings [9]) and process evolution (deep sub-micronic with thinner nitridied oxides). Single events effects (if not destructive) are not thought to be an acute problem but should justify medium term investigation as the MAPS to be used are analogue circuits that may tolerate soft errors. A modest percentage of the pixels can give erroneous data, but this proportion should be determined as well as the effects on the more modest digital part.

**ACKNOWLEDGEMENTS:**

The neutron irradiations were performed at the CERI (CNRS) at Orléans (France). The assistance of J. Briaud and co-workers (CERI, Orléans) during irradiation and for dosimetry is gratefully acknowledged. Some of the nickel samples used for dosimetry was kindly provided by an IN2P3/Orsay technical group. I particularly acknowledge the valued work of my colleagues, Y. Degerli for is involvement in the design of the chip, Y. Li for the development of some software. The contribution of our colleagues in IPHC/ULP for the Data Acquisition developments is also acknowledged.

**REFERENCES:**

[1] J. Collot et al., Nucl. Inst. and Meth. In Physics Research,A 350, 525-529 (1994)
[2] Rohn Truell, Phys. Rev., 116,890-892 (1959)
[3] N. Fourches, J. Appl. Phys. 77 (8), 3684 (1995)
[4] M. Huhtinen, Nucl. Inst. and Meth. In Physics Research A, 491, 194-215 (2004)

[5] G.C. Messenger, IEEE Transactions on Nuclear Science, Vol 39, No 3,468 (1992)
[6] N. Fourches, Doctoral Thesis, Université Louis Pasteur, Strasbourg(1989)
[7] P. Sigmund, Appl. Phys. Letters, Volume 14, Number 3, 114 (1969)
[8] H.J. Stein and R. Gereth, J. Appl. Phys. Vol 39, No 6, 2890-2904(1968)
[9] El-Sayed Eid et al., IEEE Transactions On Nuclear Science,Vol. 48,No. 6,1796(2001)